\documentstyle[aps,preprint,epsfig]{revtex}

\tightenlines

\newcommand{\be}{\begin{equation}}
\newcommand{\ee}{\end{equation}}
\newcommand{\bea}{\begin{eqnarray}}
\newcommand{\beas}{\begin{eqnarray*}}
\newcommand{\eea}{\end{eqnarray}}
\newcommand{\eeas}{\end{eqnarray*}}
\newcommand{\ba}{\begin{array}}
\newcommand{\ea}{\end{array}}
\newcommand{\del}{\partial}
\begin{document}

\preprint{\vbox{\hbox{IFT-P094/2000}}}

\title{Compton scattering in Noncommutative Space-Time at the NLC}

\author{Prakash Mathews\footnote{e-mail:mathews@ift.unesp.br}}

\address{
Instituto de Fisica Teorica, Universidade Estadual Paulista\\ 
Rua Pamplona 145, 01405-900, Sao Paulo, SP, Brazil.
        }

%\date{November, 2000}

\maketitle 

\begin{abstract} 
We study the Compton scattering in the noncommutative counter 
part of QED (NC QED).  Interactions in NC QED have momentum 
dependent phase factors and the gauge fields have Yang-Mills 
type couplings, this modifies the cross sections and are 
different from the commuting Standard Model.  Collider signals 
of noncommutative space-time are studied at the Next Linear 
Collider (NLC) operating in the $e \gamma$ mode.  Results for 
different polarised cases are presented and it is shown that
the Compton process can probe the noncommutative scale in the
range of 1 - 2.5 TeV for typical proposed NLC energies.

\end{abstract}

\vskip0.5in

%%%%%%%%%%%%%%%%%%%%%%
\section{Introduction}

	The idea that the space-time could be noncommutative (NC) dates 
back to the work of Snyder \cite{sny} in 1947.  Noncommutative field 
theory (NCFT) has been of recent interest following the realisation in 
string theory that NC spaces comes about in a specific low energy limit 
of D-branes dynamics in the presence of certain $U(1)$ constant 
background magnetic field \cite{st_nc}.  Prior to this, noncommutative 
geometry and the field theoretical construction on it was developed purely 
in general mathematical framework \cite{cons}.  The noncommutativity of 
such a space can be characterised by the commutator of coordinates $x^\mu$
\begin{equation}
[x_\mu,x_\nu]=i ~\theta_{\mu\nu} \,, 
\end{equation}
where the matrix $\theta_{\mu\nu}$ is real, antisymmetric and 
commutes with space-time coordinate $x^\mu$.  The NC parameter 
$\theta_{\mu \nu}$ has dimensions of area and reflects the 
extent to which the particular coordinates noncommute.  Further 
a NC scale $\Lambda_{NC}$ can be introduced above which the 
coordinates are noncommuting
\begin{equation}
[x_\mu,x_\nu]=i ~\frac{c_{\mu\nu}}{\Lambda^2_{NC}} \,, 
\end{equation}
where $c_{\mu\nu}$ has the same properties as $\theta_{\mu\nu}$
but the strength of noncommutativity has been absorbed in the
NC scale $\Lambda_{NC}$. 

	Various aspects of field theory on noncommutative space have 
been analysed.  The perturbative structure and the renormalisability 
of these theories have been studied \cite{renorm}.  One of the major 
out come of these studies is that the ultraviolet and infrared effects 
in noncommutative field theory do not decouple \cite{mwala}.  Hence 
physics at short distance can affect physics at long distance and vice 
versa.  The causal behaviour of space-space and space-time 
noncommutativity has been studied and shown that while space-space 
noncommutativity is causal and unitary, the space-time is acausal 
and non unitary in NCFT, but it could be restored in string theory 
\cite{causal}.  Apart from string theory \cite{causal}, unitarity 
of space-time nonlocality can also be restored in super Yang-Mills 
theory \cite{sym}.  Noncommutative QED (NC QED) has been shown
to be one loop renormalisable \cite{qed,qed1}.  
The C, P, T properties of NC QED has also been studied and is found
to be CPT invariant for space-space and space-time noncommutativity
\cite{cpt}.  
There has also been 
attempts to formulate a noncommutative Glashow-Weinberg-Salam model 
\cite{gws}.  Hence it would be interesting to explore the possible 
effects of noncommutativity at the collider level.  

Noncommutative effects at the colliders has been studied for the 
first time by Hewett {\it et.~al.} \cite{HR}.  They have considered 
several $2 \rightarrow 2$ processes in $e^+ e^-$ collisions at the 
Next Linear Collider (NLC) and shown that NC scale of the order of 
a TeV can be probed at the NLC.  
Though the NC scale $\Lambda_{NC}$ could be the Planck or the GUT 
scale it was argued in the context of the recent work on possible 
TeV scale quantum gravity \cite{add,rs} that one could as well 
consider the NC scale $\Lambda_{NC}$ to be not too far from the 
TeV scale \cite{HR}.  They have considered the Moller, Bhabha and 
annihilation process in the $e^+ e^-$ mode and $\gamma \gamma
\rightarrow \gamma \gamma$ process in the $\gamma \gamma$ mode of 
NLC.  The 95 \% lower bound on the $\Lambda_{NC}$ for 
the various process obtained are (a) 1700 GeV for Moller, (b) 1050 
GeV for Bhabha, (c) in the range 740-840 GeV for annihilation and 
(e) for $\gamma \gamma \rightarrow \gamma \gamma$ the space-time 
NC yields 700-800 GeV while space-space NC gives 500 GeV \cite{HR}.
Other phenomenological aspects of noncommutativity has been studied
\cite{atom,sheik,lamp} and bounds on the $\theta$ parameter obtained
for high and low energy limits.

The NLC is  planned to operate in $e^+ e^-$, $\gamma \gamma$ 
and $e \gamma$ mode.  At high energy and luminosity, the $e^+ e^-$ 
collider can yield a $\gamma$ beam of comparable energy and luminosity 
using the laser back scattering technique \cite{ginz}.  The NLC is 
an ideal testing ground of the Standard Model (SM) and a very 
effective probe of 
possible physics that may lie beyond the SM.  The experiment at 
the NLC also provides a great degree of precision because of the 
relatively clean initial states and the degree of precision can be 
enhanced by using the polarisation of the initial states.  Hence we 
have considered the possibility of testing the NC effects at the NLC 
in the $e \gamma$ mode by studying the Compton scattering.  

In sec.~2 we introduce the basics of NC QED.  Sec.~3 we present
the cross section for Compton scattering in NC space.  The numerical
results and possible reach of the NLC is discussed in sec.~4.  
Finally we summarise our results in sec.~5.

%%%%%%%%%%%%%%%%
\section{NC QED}

	The one loop UV divergent structure of U(1) Yang-Mills 
theory on noncommutative $R^4$ has been analysed \cite{qed}.
The mathematical structure of the quantum field theory over a 
noncommutative space is not yet well understood but by 
assuming that the quantum theory is defined by the generating 
functional of the theory, it was shown that the theory is 
renormalisable at the one loop level \cite{qed}.  The matter 
fields where introduced and the perturbative aspects of 
NC QED have been analysed \cite{qed1}.  The NC QED action is given by
\begin{eqnarray}
S_{NC~QED}=\int d^d x \left( -\frac{1}{4 g^2} F_{\mu\nu} * F^{\mu\nu}
+ \bar \psi * \gamma^\mu i D_\mu \psi -m \bar \psi * \psi \right ) \,,
\label{act}
\end{eqnarray}
where the $*$-product is defined as
\begin{eqnarray}
A*B(x) = \exp\left(\frac{i}{2} \theta_{\mu\nu} \frac{\del}{\del \xi^\mu}
\frac{\del}{\del \eta^\nu}\right) A(x+\xi)B(x+\eta)|_{\xi,\eta=0} 
\,.
\end{eqnarray}
The noncommutativity of space modifies the algebra of functions
and even in the $U(1)$ case the field strength is nonlinear and has
the form $F_{\mu\nu} = \del_\mu A_\nu - \del_\nu A_\mu - i 
[A_\mu,A_\nu]_M$.  The Moyal bracket $[A,B]_M$ is a commutator under 
$*$-product.  The covariant derivative for the matter fields is given 
by
\begin{eqnarray}
D_\mu \psi = \del_\mu \psi - i ~A_\mu * \psi \,, \quad \quad 
D_\mu \bar \psi = \del_\mu \bar \psi + i ~\bar \psi * A_\mu \,.
\end{eqnarray}
There exist other possible choice of covariant derivative which are
related to the above by charge conjugation and are detailed in \cite{cpt}.
The action eq.~(\ref{act}) is invariant under the NC gauge transformation
\begin{eqnarray}
A_\mu(x) \rightarrow A^\prime(x) &=& U(x)*A_\mu(x)*U^{-1}(x)
+i U(x)*\del_\mu U^{-1}(x) \,,
\nonumber\\
\psi(x) \rightarrow \psi^\prime (x) &=& U(x) * \psi (x) \,,
\quad \quad
\bar \psi(x) \rightarrow \bar \psi^\prime (x) = \bar 
\psi (x) * U(x)^{-1} \,,
\end{eqnarray}
where $U(x)=\exp(i \alpha(x))_*$ is defined as an infinite series
of the scalar function $\alpha(x)$ under $*$-product and $U(x)^
{-1}$ is its inverse.  Note that due to the $*$-product 
algebra of the NC space, the Lagrangian is no longer
invariant under the NC gauge transformation but the 
action is.  Hence this gives the basic ingredients needed to derive the 
Feynman rules \cite{qed1}.  Each of the interaction vertex picks 
up a phase factor which depends on the momentum and even in the 
$U(1)$ case, $A_\mu$ couples to itself and there exist three and 
four point photon vertices.  The propagators are the same as the 
commuting theory.   Due to noncommutativity, the theory is 
manifestly nonlocal in the noncommuting coordinates and this 
leads to violation of Lorentz invariance.  

%%%%%%%%%%%%%%%%%%%%%%%%%%%%%%%%%%%%%%
\section{Compton scattering in NC QED}

	The Feynman diagrams that contributes to the tree level process
$\gamma(k_1) e(p_1) \rightarrow \gamma(k_2) e(p_2)$ in the NC QED has 
the usual $s$- and $u$-channel diagrams and in addition there is also a 
$t$-channel diagram coming from the triple photon vertex.  The 
amplitude is given by
\begin{eqnarray}
{\cal M}&=&-i~ g^2 ~\exp(\frac{i}{2} p_1 \wedge p_2 )
~{\epsilon_1}_\mu(k_1) ~{{\epsilon_2}^*_\nu} (k_2)
\nonumber \\
&& \left [
   \left  (
\frac{1}{\hat s} 
\bar u(p_2) \gamma_\nu (\not p_1 + \not k_1) \gamma_\mu u(p_1)
+ \frac{V^{\mu \nu \rho} }{\hat t} \bar u(p_2) \gamma_\rho u(p_1) 
   \right ) \exp \left( \frac{i}{2} ~p_{12} \wedge k_2 \right) 
\right . \nonumber \\
&& \left .
     +\left (
\frac{1}{\hat u} 
\bar u(p_2) \gamma_\mu (\not p_1 - \not k_2) \gamma_\nu u(p_1) 
- \frac{V^{\mu \nu \rho} }{\hat t} \bar u(p_2) \gamma_\rho u(p_1) 
      \right ) \exp \left( \frac{-i}{2} ~p_{12} \wedge k_2 \right) 
\right] \,,
\end{eqnarray}
where $g$ is the NC coupling, $p_{12}=p_1-p_2$, the $hat$ on the
Mandelstam variables corresponds to the $e \gamma$ subprocess
and the three photon vertex function $V$ is given by
\be
V_{\mu\nu\rho} =  g_{\mu\rho} (2 k_1 - k_2)_\nu
                + g_{\nu\rho} (2 k_2 - k_1)_\mu
                - g_{\mu\nu}  (  k_1 + k_2)_\rho \,.
\ee
The $\wedge$-product is defined as
\begin{equation}
p_1 \wedge p_2 = \theta_{\mu \nu} ~p_1^\mu ~p_2^\nu
=
\frac{c_{\mu \nu}} {\Lambda^2_{NC}} ~p_1^\mu ~p_2^\nu \,.
\end{equation}
The NC effects appear as the phase factors.  In the commuting limit 
$\theta \rightarrow 0$ the SM diagrams are 
recovered.  Due to the triple photon NC contribution, care should be 
taken to check the Ward identity and to retain only the physical 
photon degrees of freedom.  The NC effects at the cross section 
could arise from the interference of diagrams which could pick up 
a phase or from the explicit three and four photon vertex diagrams.  

At the NLC, $\gamma$ beam can be obtained from laser back 
scattering and hence has a distribution in energy and helicity 
of the parent electron and laser beam.  The differential cross 
section for the Compton scattering at the NLC is given by
\begin{eqnarray}
\frac{d \sigma}{d \Omega} = \alpha^2
\int d x \frac{f(x)}{x s}
\left (
 \frac{1+P_{e_2} \xi_2(x)}{2} |{\cal M}_{++}|^2
+\frac{1-P_{e_2} \xi_2(x)}{2} |{\cal M}_{+-}|^2
\right ) \,,
\label{cs}
\end{eqnarray}
where $\alpha=g^2/4 \pi$, $x$ is the fraction of the parent electron 
energy carried by the photon, $s$ corresponds to the $cms$ lab frame 
of the $e^+ e^-$ pair and $P_{e_2}$ is the beam polarisation of the 
electron beam.  The photon number density $f(x)$, the Stokes parameter 
$\xi_2(x)$ are functions of $x$, the parent electron and laser beam 
polarisation.  The details are given in the Appendix.  
It has been suggested in \cite{lamp} to study the low energy Thomson 
limit of the Compton scattering in NC case to find the physical value 
of NC coupling as in the QED case \cite{gold}.  Whether this property 
of Thomson limit holds for the NC case is being investigated \cite{lamp}.

The matrix 
element square $|M_{ij}|^2$ corresponds to the subprocess $\gamma(i) 
~e (j) \rightarrow \gamma ~e$, where i, j denotes the respective initial 
helicities while the final helicities are summed over.  The polarised
$|M_{ij}|^2$ are given by
\begin{eqnarray}
|M_{++}|^2 &=& -\frac{\hat s}{\hat u} 
+ 4 ~\frac{\hat s^2}{\hat t^2} 
~\sin^2 \Delta_C \,,
\nonumber \\
|M_{+-}|^2 &=& -\frac{\hat u}{\hat s} 
+ 4 ~\frac{\hat u^2}{\hat t^2} 
~\sin^2 \Delta_C \,,
\label{me}
\end{eqnarray}
where the noncommuting phase is 
\be
\Delta_C=\frac{1}{2} ~p_{12} \wedge k_2 \,.
\ee
The NC effects for the Compton scattering reside in the even function 
$\sin^2 \Delta_C$ and taking the commuting limit $\Delta_C \rightarrow 
0$ one recovers the SM form.  The commuting limit seems possible at the 
tree level but in the far infrared the physics might look different
due to loop effects and ultraviolet/infrared mixing \cite{qed,qed1}.  
To perform the phenomenology we assume the terms independent of
the NC scale as the SM contribution and study the effects induced
by the noncommutativity.  

The NC Compton phase in the above equation is an out come of a 
combination of phase effects arising from the $t$-channel triple 
photon vertex diagram, its interference with the $s$- and $u$-channel 
diagrams and the interference between the $s$- and $u$-channel diagrams.  
The polar angle dependence of the above equation is different for the 
commuting SM and can be suitably exploited to distinguish
the NC QED contribution.  The cross section eq.~(\ref{cs}) is a function 
of the polarisation of the electron, the parent electron and laser 
beams.  The polarisation can be used as a useful tool for the analysis.  

To evaluate $\Delta_C$ we have to go to a specific frame and we choose 
the $e \gamma$ $cms$ frame 
\begin{eqnarray}
k_1&=& \frac{\sqrt{\hat s}} {2} (1, ~0, ~0, ~1) \,, \qquad
~~k_2= \frac{\sqrt{\hat s}} {2} (1, ~\sin \theta \cos \phi, 
   ~\sin \theta \sin \phi, ~\cos\theta) \,,
\nonumber \\
p_1&=& \frac{\sqrt{\hat s}} {2} (1, ~0, ~0, -1) \,, \qquad
p_2= \frac{\sqrt{\hat s}} {2} (1, -\sin \theta \cos \phi, 
   -\sin \theta \sin \phi, -\cos\theta) \,,
\end{eqnarray}
where $\theta$ is the polar angle in the $e \gamma$ $cms$ 
frame and the corresponding Mandelstam variables are 
$\hat t=(k_1-k_2)^2 =-\hat s (1-\cos \theta)/2$,
$\hat u=(k_1-p_2)^2=-\hat s (1+\cos \theta)/2$ and 
$\hat s=x~s$.  The Compton phase factor $\Delta_C$ 
in this frame has the form
\begin{eqnarray}
\Delta_C= \frac{\hat s}{8 \Lambda^2_{NC}}
( &-& c_{13} \sin {\theta} \cos {\phi} 
   -  c_{23} \sin {\theta} \sin {\phi} 
\nonumber \\
  &+& c_{01} \sin {\theta} \cos {\phi} 
   +  c_{02} \sin {\theta} \sin {\phi} 
   +  c_{03} (-1+\cos {\theta}) 
) \,.
\end{eqnarray}
Note that $\Delta_C$ gets contributions from space-space and 
space-time noncommutativity.  From the phase factor there is 
an additional polar angle dependence in the cross section and 
unlike the usual $2 \rightarrow 2$ process there is also an 
azimuthal angle dependence.  The explicit form of the phase 
factor depends on the amount of noncommutativity of the 
particular noncommuting coordinates, ie.~the magnitude of the 
components of $c_{\mu\nu}$.  For the numerical analysis we choose 
a few specific cases described in the next section.

%%%%%%%%%%%%%%%%%%%%%%%%%%%
\section{Numerical Results}

	Before we present our numerical analysis it is important to
point out that the phase factor $\Delta_C$ depends on the choice of 
frame and hence the cross section is not Lorentz invariant.  But 
nonlocality and violation of Lorentz invariance are manifestation
of NC spaces and if the space is indeed noncommuting above some scale
$\Lambda_{NC}$ the effects of these violation could be observed close 
to $\Lambda_{NC}$.  This could for instance be an azimuthal angle
dependence in a $2 \rightarrow 2$ process.  As was pointed out in 
\cite{HR}, two experiments could still compare results by converting 
to a common frame of reference which could be some slowly varying 
astronomical co-ordinate system.  

To perform the analysis we consider two cases {\it viz.} ({\sf I}) 
$c_{0i}=0$ and $c_{ij} \not =0$ (space-space NC) and ({\sf II}) 
$c_{0i}\not =0$ and $c_{ij}=0$ (space-time NC).   Further assuming 
the components of $c_{\mu\nu}$ to be of order of unity 
\begin{eqnarray}
\Delta_C^{\sf I}  &=& - \frac{\hat s}{8 \Lambda_{NC}^2} 
                      \sin \theta \cos (\phi-\beta) \,;
\qquad \qquad \qquad 
\qquad \qquad \qquad \quad ~~
c_{0i}=0 \,, ~~~~~ c_{ij} \not =0 \,, 
\nonumber \\
\Delta_C^{\sf II} &=&   \frac{\hat s}{8 \Lambda_{NC}^2} 
(- \cos \alpha 
 + \cos \alpha \cos \theta 
 + \sin \alpha \sin \theta \cos (\phi-\beta) \,;
\qquad  
c_{0i} \not =0 \,, ~~~~~ c_{ij} =0 \,,
\end{eqnarray}
where in case-{\sf I} the noncommutativity is parametrised as $c_{13}= 
\cos \beta$, $c_{23}= \sin \beta$ and in case-{\sf II}, $c_{01}=\sin 
\alpha \cos \beta$, $c_{02}= \sin \alpha \sin \beta$ and $c_{03}=\cos 
\alpha$.  The angle $\beta$ fixes the origin from where the azimuthal 
angle $\phi$ is measured and is chosen to be $\beta=\pi/2$.  This 
chooses a specific direction and hence a violation of the Lorentz 
symmetry.  It is instructive to look at the explicit form of the the 
phase factor for $\beta=\pi/2$ and for the various values of $\alpha$
\begin{eqnarray}
\Delta_C^{\sf I}  &=& - \frac{\hat s}{8 \Lambda_{NC}^2} 
                      \sin \theta \sin \phi \,;
~~~\qquad \qquad \qquad \qquad
c_{13}=0, ~~~~ c_{23}=1 \,,
\nonumber \\
\Delta_C^{\sf II}|_{0} &=&   \frac{\hat s}{8 \Lambda_{NC}^2} 
(- 1 + \cos \theta)  \,;
~~\qquad \qquad \qquad \qquad
c_{01}=0, ~~~~ c_{02}=0, ~~~~~~~ c_{03}=1 \,,
\nonumber \\
\Delta_C^{\sf II}|_{\frac{\pi}{4}} &=&  \frac{\hat s}{8\sqrt{2} 
\Lambda_{NC}^2} (- 1 + \cos \theta + \sin \theta \sin \phi) \,;
\qquad
c_{01}=0, ~~~~ c_{02}=\frac{1}{\sqrt{2}}, 
~~~~ c_{03}=\frac{1}{\sqrt{2}} \,,
\nonumber \\
\Delta_C^{\sf II}|_{\frac{\pi}{2}} &=&  \frac{\hat s}{8 
\Lambda_{NC}^2} \sin \theta \sin \phi \,;
~~~~~\qquad \qquad \qquad \qquad
c_{01}=0, ~~~~ c_{02}=1, ~~~~~~~ c_{03}=0 \,.
\label{phc}
\end{eqnarray}
The above cases have definite meaning with respect to the constant 
background field responsible for the noncommutativity \cite{st_nc}.  
The space-space noncommutativity $c_{ij}\neq 0$ corresponds to a 
background magnetic field in the direction perpendicular to $ij$ 
while the space-time noncommutativity $c_{0i}\neq 0$ corresponds 
to a background
electric field in the direction characterised by an angle $\alpha$ 
with respect to the beam direction.  Case-{\sf I} and case-{\sf IIc} 
($\alpha=\pi/2$) have the same form at the cross section level.  
Case-{\sf IIa} does not have an azimuthal angle dependence.  For 
the numerical analysis we will only consider the distinct cases-{\sf 
I}, {\sf IIa} and {\sf IIb} ($\alpha=\pi/4$).  We have taken $x$ 
in the range [0.1,0.82] and have a cut on the  polar angle in the 
$e^+ e^-$ $cms$ lab frame, $|\cos \theta_{lab}| < 0.95$.  The 
results are presented for $\sqrt{s}$ in the range [500,1500] GeV.  
Further the various polarisation states are considered denoted by 
$(P_{e1}, P_{l1},P_{e2})$ where $P_{e1}$ and $P_{l1}$ corresponds 
to the parent electron and laser beam and $P_{e2}$ the other 
electron beam which interacts with the photon beam in the subprocess.  
The polarisation state $(0,0,0)$ corresponds to the unpolarised case. 

\begin{figure}
\centerline{\mbox{
\epsfig{file=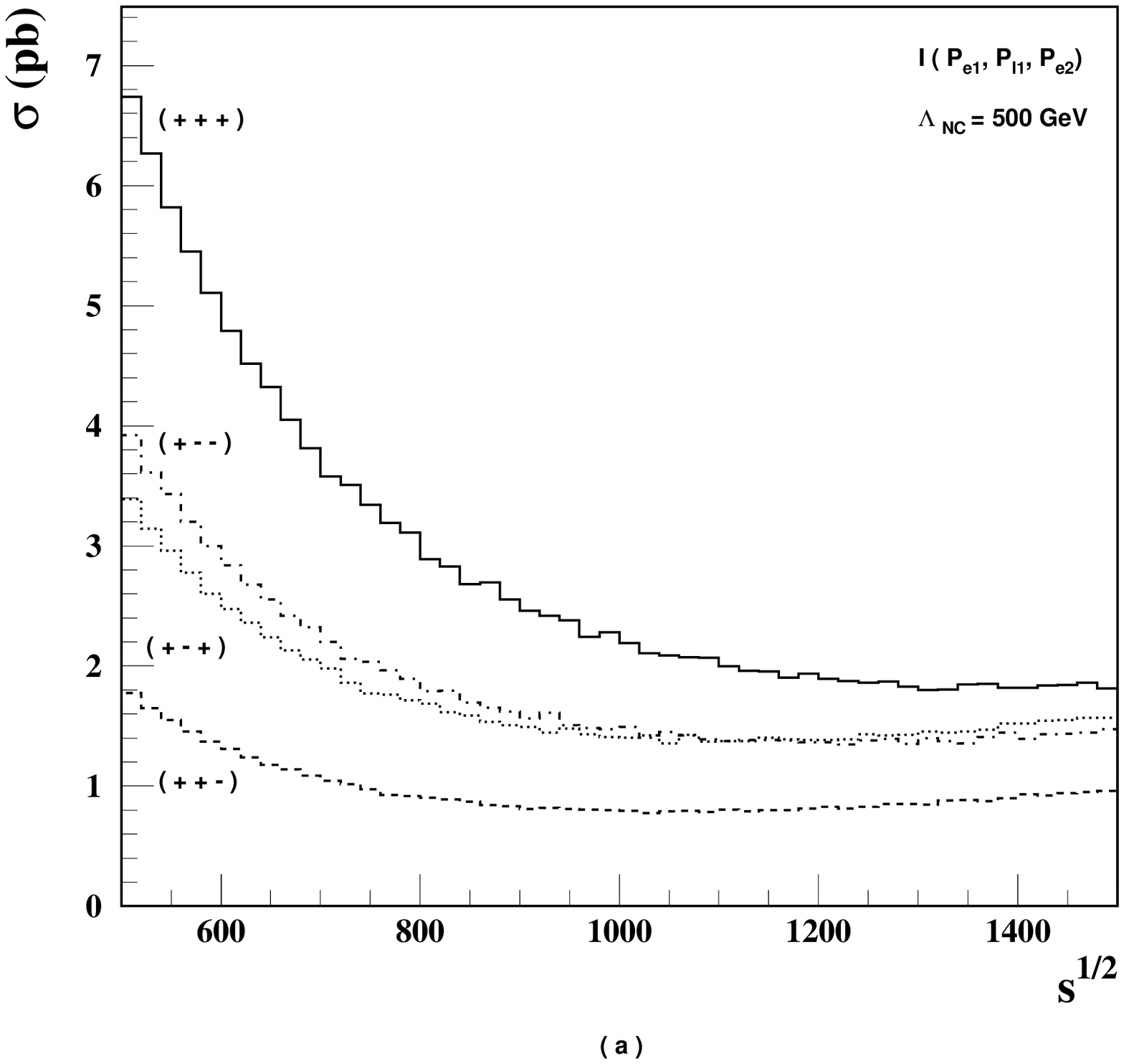,width=0.56\textwidth}
\epsfig{file=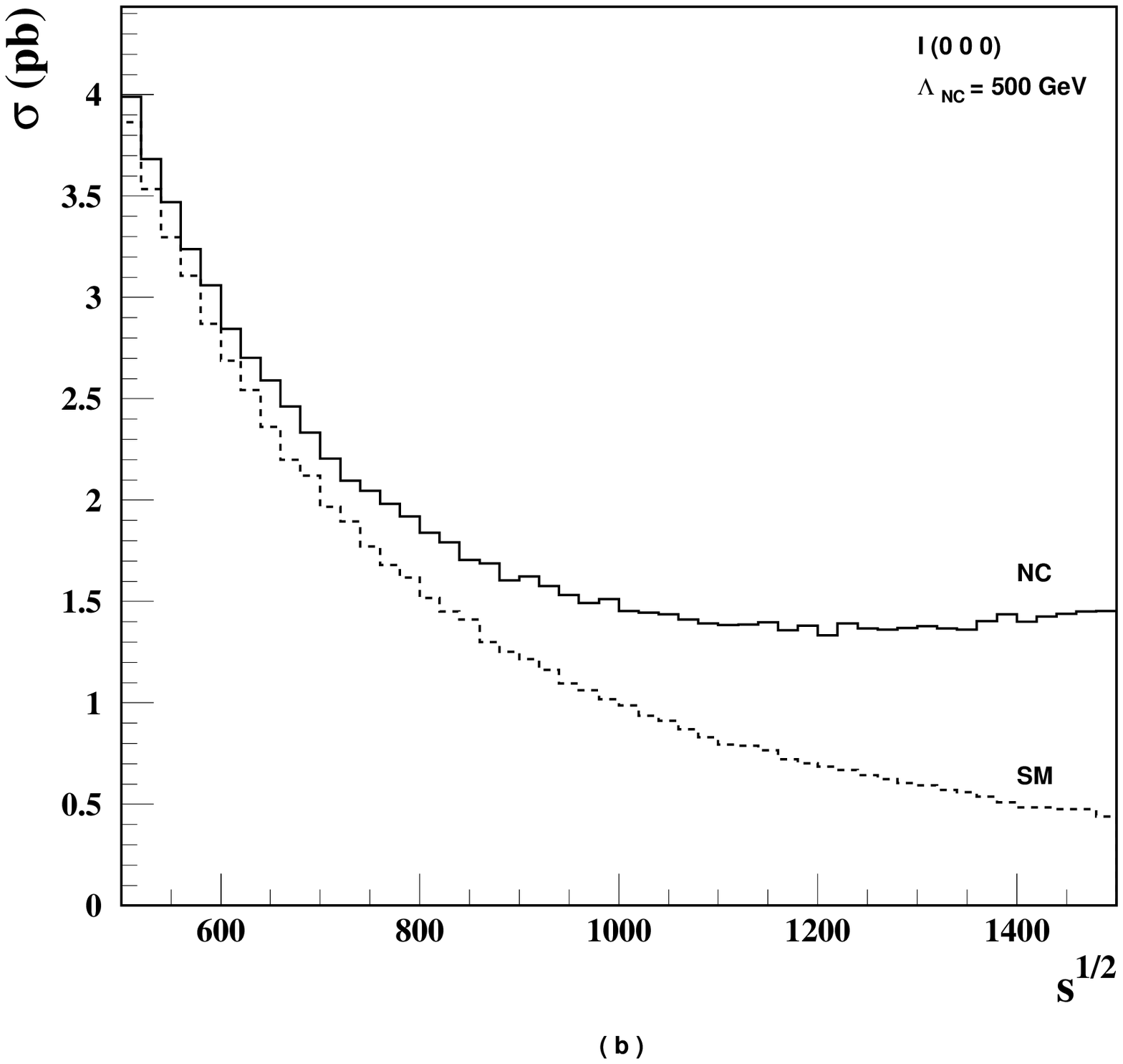,width=0.56\textwidth}
}} 
\caption{
(a) Total cross section as a function of $\sqrt{s}$ for 
various polarisation states $(P_{e1},P_{l1},P_{e2})$. 
(b) Unpolarised total cross section $(0,0,0)$ as a function 
of $\sqrt{s}$ for the NC and SM.
}
\label{roots}
\end{figure}

In fig.~\ref{roots} the various polarisation states of case-{\sf I} 
has been plotted as a function of $\sqrt{s}$ in the range [500,1500] 
GeV for $\Lambda_{NC}=500$ GeV.  The polarisation state $(+,+,+)$ is 
the most dominant.  To compare this with the commuting limit (SM), 
we have plotted the unpolarised cross section $(0,0,0)$ for NC and SM 
for case-{\sf I} with $\Lambda_{NC}=500$ GeV.  The SM contribution 
drops much faster then the NC contribution at larger $\sqrt{s}$ and 
hence can be probed at the NLC.  This behaviour is typical for each 
of the NC cases under consideration and as the NC scale $\Lambda_
{NC}$ is increased the drop with $\sqrt{s}$  becomes similar to the 
SM case though it is still higher.  To study the reach possibilities 
at NLC as a function of the NC scale $\Lambda_{NC}$ we perform a 
$\chi^2 (\Lambda_{NC})$ fit using
\be
\chi^2 (\Lambda_{NC}) = \frac{L}{\sigma_{SM}} (\sigma_{SM} - 
                        \sigma_{NC}(\Lambda_{NC}))^2 \,,
\ee
where $L$ is the luminosity, $\sigma_{SM}$ is the SM total cross 
section and $\sigma_{NC}$ is the NC QED total cross section.  We 
choose an integrated luminosity $\int L=500$ fb$^{-1}$ for the 
calculation.  We obtain the 95 \% CL lower bound on $\Lambda_{NC}$ 
by demanding $\chi^2(\Lambda_ {NC})\geq 4$.  The reach at NLC for 
various NC cases have been tabulated in Table \ref{tab1} for $\sqrt
{s}=$ 500, 1000 and 1500 GeV.  The various polarisation states
give different search limits and the range has been summarised in 
Table \ref{tab1}.

\begin{figure}
\centerline{\mbox{
\epsfig{file=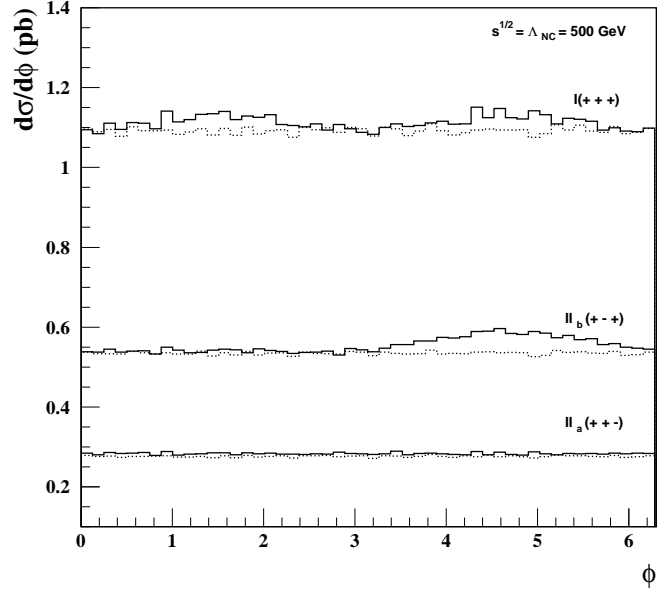,width=0.6\textwidth}
}} 
\caption{
Azimuthal angle dependence for various case of noncommutativity
and polarisation.  The solid line denotes NC and the doted line 
denotes the corresponding SM.  
}
\label{phi}
\end{figure}

	The the typical azimuthal angle dependence of the various NC
cases have been plotted in fig.~\ref{phi}, for $\sqrt{s}=\Lambda_{NC}
=500$ GeV.  Inspection of eq.~(\ref{phc}) shows the $\phi$ dependence, 
which corresponds to the choice of noncommutativity.  Case-{\sf IIa} is 
independent of $\phi$ while Case-{\sf I} and Case-{\sf IIb} have distinct 
$\phi$ dependence as shown in fig.~\ref{phi}.  

\begin{figure}
\centerline{\mbox{
\epsfig{file=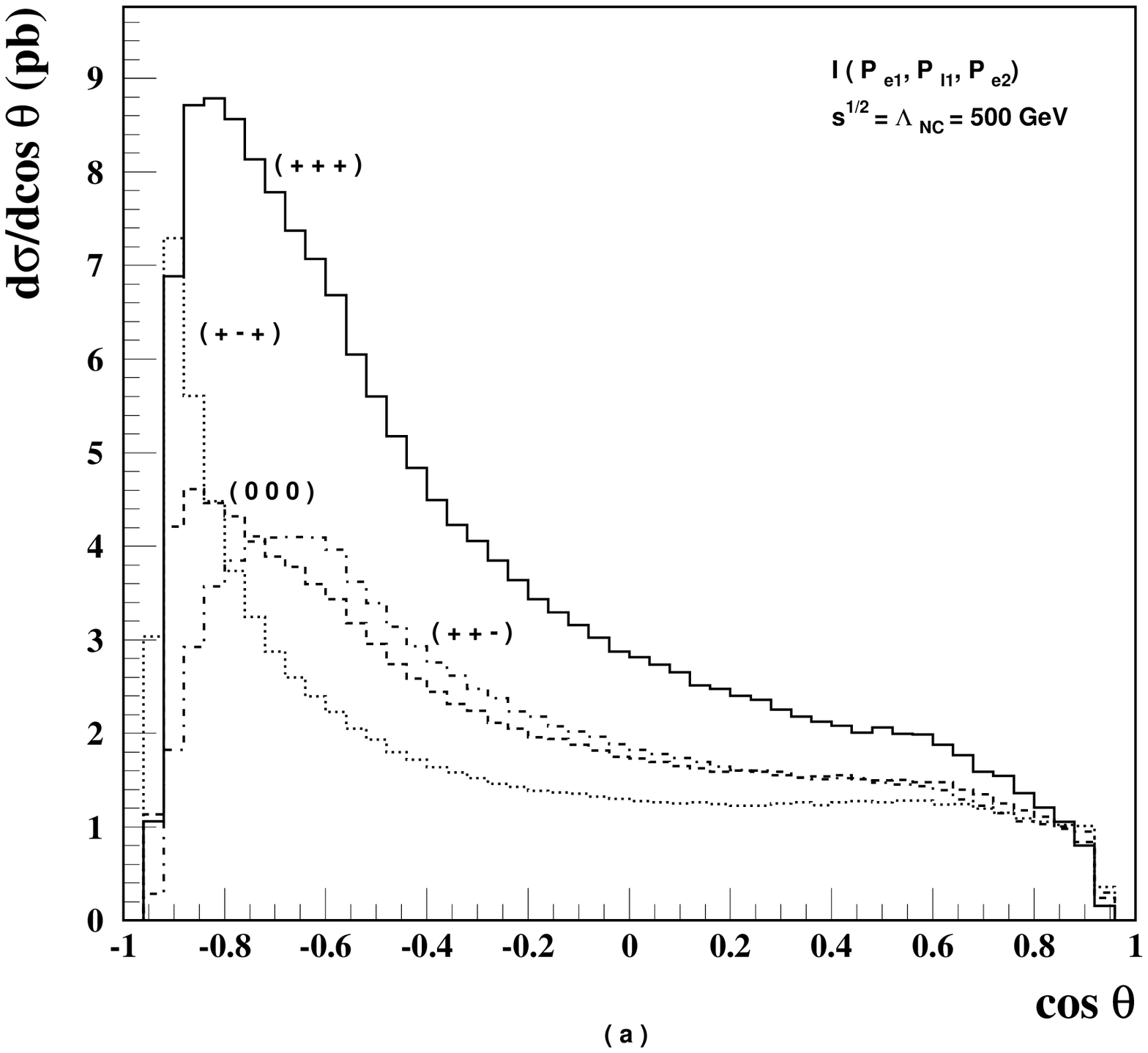,width=0.55\textwidth}
\epsfig{file=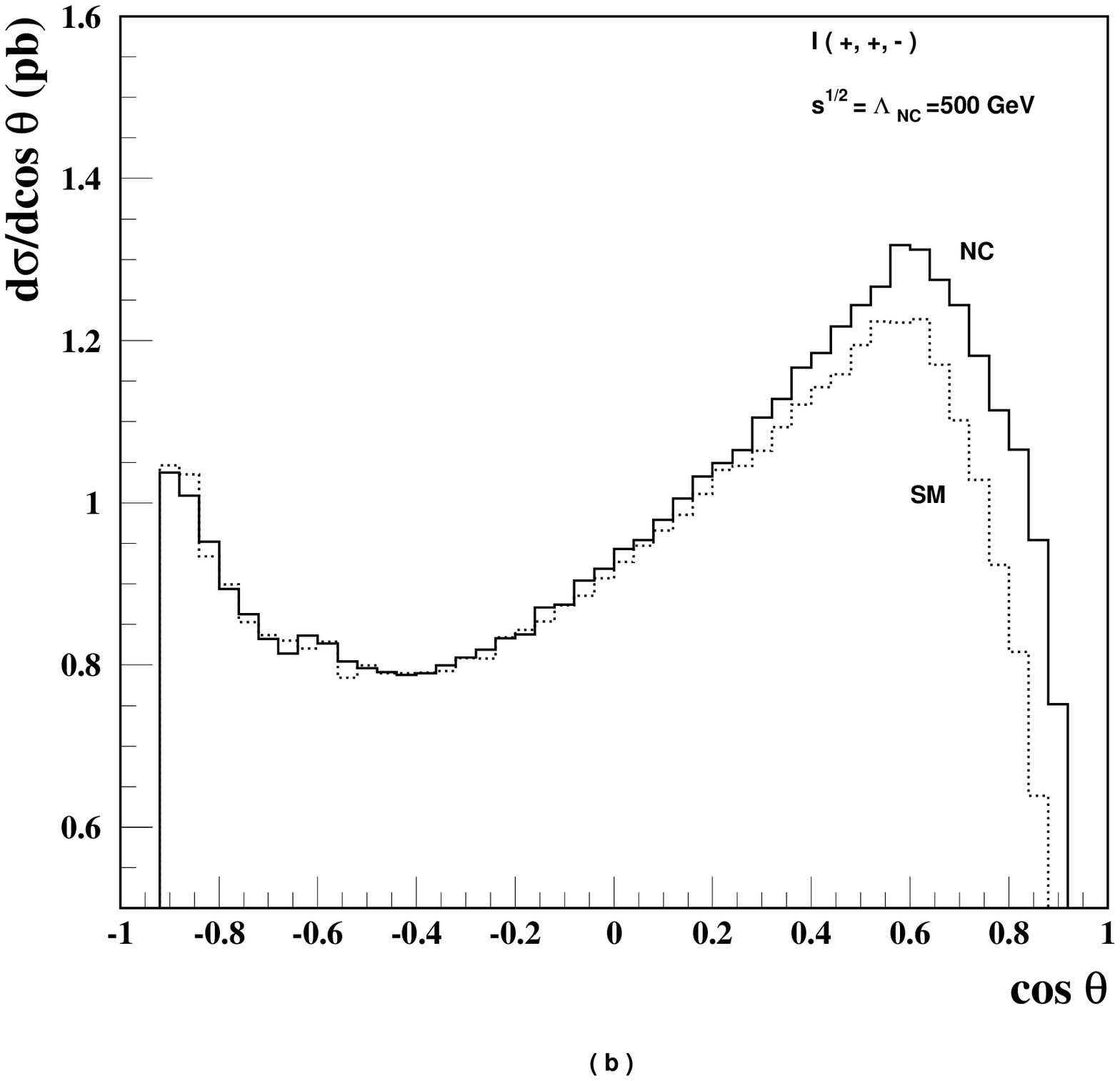,width=0.55\textwidth}
}}
\caption{
(a) The differential cross section for noncommutativity case-{\sf I} 
as a function of $e \gamma$ $cms$ polar angle $\cos \theta$ for $\sqrt{s}
=\Lambda_{NC}= 500$ GeV for various polarisation states.
(b) The differential cross section for the polarisation state $(+,+,-)$
for NC and SM.  
}
\label{cos}
\end{figure}

	The polar angle dependence of the SM and NC is different 
as can be see in eq.~(\ref{me}).  The SM has $1/u$ dependence and hence 
will peak in the backward direction while the NC case has an additional
$1/t^2$ dependence and hence also has a forward dominance.  The NC cross 
section has been plotted as a function of $e \gamma$ $cms$ polar angle 
$\cos \theta$ for $\sqrt{s}=\Lambda_{NC}= 500$ GeV.  In fig.~\ref{cos}a
the various polarisation states have been compared.  To exhibit the 
forward dominance of NC case we have plotted in fig.~\ref{cos}b the $(+,+,-)$ 
polarisation state in comparison to the SM.  

%%%%%%%%%%%%%%%%%%%%%
\section{conclusions}

In this paper we have considered the Compton scattering in the
noncommutative QED and studied the various distributions at the
NLC in the $e \gamma$ mode.  The NC phase factor $\Delta_C$ has 
contributions from space-space and space-time noncommutativity.
We have considered the two cases separately.  The various
noncommutative cases considered have distinct azimuthal angle 
dependence.  The polar angle dependence for noncommuting and 
the commuting SM is also different and can be used to probe the
noncommutativity.  The 95\% CL reach possibilities of NC scale
$\Lambda_{NC}$ for various noncommuting cases at the NLC operating
in the $\sqrt s$ range [500,1500] GeV has been given in 
Table.~\ref{tab1}.
 
%%%%%%%%%%%%%%%%%%%%%%%%
\acknowledgements   

I would like to thank Keshav Dasgupta and Oscar J.~P.~Eboli for useful 
discussions.  I also thank Oscar for going through the manuscript and 
for useful comments.  Thanks are also due to Adi Armoni for his comments
on the manuscript.  The referee is also thanked for useful comments.
This work was supported by FAPESP (Processo: 99/05310-9).

%%%%%%%%%%%
\appendix
\section*{}

This appendix contains the useful formulae for the distribution of
photon produced using the laser back scattering technique.
The electron beam responsible for the production of photon beam 
is called the parent electron.  For further details regarding 
laser back scattering technique, refer to the original work 
\cite{ginz}. The photon beam produced by laser back 
scattering has number density given by the distribution
\begin{eqnarray}
f(x,P_e,P_l;z)=\frac{2 \pi \alpha^2}{m_e^2 z \sigma_C} C(x) \,,
\end{eqnarray}
where $P_e$ is the polarisation of the electron beam and is taken at 
90\% level $(P_e =\pm 0.9)$, $P_l$ is the laser beam polarisation and can 
achieve almost 100 \%  $(P_l=\pm 1)$.  The fraction of the parent 
electron beam energy carried by the photon is denoted by $x$.  The
maximum value $x$ can take is $0.82$, above this value the laser 
beam and back scattered photon can pair produce $e^+ e^-$ pairs and 
hence decreases the efficient of the process.  $x_{max}=z/(1+z)$ and 
the optimal value for $z$ is $z_{OPT}=2(1+\sqrt{2})$.  The function 
$C(x)$ and $\sigma_C$ is given by
\begin{eqnarray}
C(x) &=& \frac{1}{1-x}+1-x-4 r (1-r)-P_e ~P_l ~r z (2r-1) (2-x) \,,
\nonumber\\
\sigma_C &=& \frac{2 \pi \alpha^2}{m_e^2 z }
\left [
\left( 1-\frac{4}{z}-\frac{8}{z^2} \right)\ln (1+z) + \frac{1}{2}
+ \frac{8}{z} - \frac{1}{2(1+z)^2}
   \right.
\nonumber\\
&& + P_e ~P_l
\left.
\left(
\left( 1+\frac{2}{z} \right)\ln (1+z)- \frac{5}{2} + \frac{1}{1+z} 
- \frac{1}{2(1+z)^2} 
\right)
  \right ] \,,
\end{eqnarray}
where $r=x/(z(1-x))$.  The Stokes parameter is defined as
\begin{eqnarray}
\xi_2(x,P_e,P_l;z)=
\frac{P_e}{C(x)}
\left(\frac{x}{1-x}+x(2r-1)^2\right)-
\frac{P_l}{C(x)} (2 r-1) \left(1-x+\frac{1}{1-x}\right) \,.
\end{eqnarray}

%%%%%%%%%%%%%%%%%%--- References
%%%%%%%%%%%%%%%%%%%%%%%%%%%%%%%%%%%%%%%%%%%%%%%%%%%%%%%
\def\MPL  {Mod. Phys. Lett.~    }
\def\NPB  {Nucl. Phys.~         }
\def\PLB  {Phys. Lett.~         }
\def\PR   {Phys. Rep.~          }
\def\PRD  {Phys. Rev.~          }
\def\PRL  {Phys. Rev. Lett.~    }
\def\RMP  {Rev. Mod. Phys.~     }
\def\NIM  {Nuc. Inst. Meth.~    }
\def\ZPC  {Z. Phys.~            }
\def\EJPC {E. Phys. J.~         }
\def\IJMP {Int. J. Mod. Phys.~  }
\def\JHEP {J. High Energy Phys.~}

\begin{table}
\begin{tabular}{|l|c|c|c|} 
%\hline 
NC       & $\sqrt{s}=500$ & $\sqrt{s}=1000$  & $\sqrt{s}=1500$   \\ 
\hline
Case-{\sf I}   & $1050(+-+)$ - $1190(+++)$    
               & $1780(+--)$ - $1840(+-+)$      
               & $2410(+++)$ - $2530(+-+)$ 
\\ \hline 
Case-{\sf IIa} & $ 890(++-)$ - $1140(+++)$    
               & $1530(++-)$ - $1730(+++)$      
               & $2080(++-)$ - $2340(+-+)$ 
\\ \hline 
Case-{\sf IIb} & $ 990(++-)$ - $1150(+++)$    
               & $1670(+--)$ - $1780(+++)$      
               & $2270(+--)$ - $2440(+-+)$ \\
%\hline 
\end{tabular}
\vspace{1cm}
\caption{95\% C.L. limits on the NC scale $\Lambda_{NC}$ in GeV 
for $\sqrt{s}=$500, 1000, 1500 GeV obtained using an integrated 
luminosity $\int L=500 ~fb^{-1}$.  The various polarisation states 
that yield the limits is indicated as $(P_{e1},P_{l1},P_{e2})$.} 
\label{tab1}
\end{table}



\begin{thebibliography}{99}

\bibitem{sny}
H. S. Snyder, Phys. Rev. {\bf 71}, 38, (1947);
{\it ibid}, Phys. Rev. {\bf 72}, 68, (1947).

\bibitem{st_nc}
A. Connes, M.R. Douglas and A. Schwarz, \JHEP {\bf 9802}, 003, (1998), 
hep-th/9711162;
M.R. Douglas and C. Hull, \JHEP {\bf 9802}, 008, (1998), hep-th/9711165;
V. Schomerus, \JHEP {\bf 9906}, 030, (1999), hep-th/9903205;
N. Seiberg and E. Witten, \JHEP {\bf 9909}, 032, (1999), hep-th/9908142.

\bibitem{cons}
A. Connes, {\it Non-commutative Geometry}, Academic Press, (1994).

\bibitem{renorm}
T.~Filk, \PLB {\bf B376}, 53, (1996);
J.C.~Varilly and J.M.~Gracia-Bondia, \IJMP {\bf A14}, 1305, (1999), 
hep-th/9804001;
M.~Chaichian, A.~Demichev and P.~Presnajder, hep-th/9812180;  
{\it ibid} hep-th/9904132;
M.~M.~Sheikh-Jabbari, \JHEP {\bf 06}, 015, (1999), 
hep-th/9903107;
C.P.Martin, D. Sanchez-Ruiz, \PRL {\bf 83}, 476, (1999)
hep-th/9903077;
T.~Krajewski, R. Wulkenhaar, hep-th/9903187;
S.~Cho, R.~Hinterding, J.~Madore and H.~Steinacker, hep-th/9903239;
E.~Hawkins, hep-th/9908052;
D.~Bigatti and L. Susskind, hep-th/9908056;
N.~Ishibashi, S. Iso, H. Kawai and Y. Kitazawa, hep-th/9910004;
I.~Chepelev and  R. Roiban, hep-th/9911098;
H.~Benaoum, hep-th/9912036;
A.~Armoni, hep-th/0005208.

\bibitem{mwala}
S. Minwalla, M, Van Raamsdonk and N. Seiberg, hep-th/9912072.

\bibitem{causal}
N. Seiberg, L Susskind and Nicolaos, \JHEP {\bf 06}, 004, (2000).

\bibitem{sym}
R.-G. Cai and N. Ohta, hep-th/0008119.

\bibitem{qed}
C.P.Martin, D. Sanchez-Ruiz, \PRL {\bf 83}, 476, (1999),
hep-th/9903077.

\bibitem{qed1}
M.~Hayakawa, \PLB {\bf B478}, 394 (2000), hep-th/9912094;
{\it ibid}, hep-th/9912167.

\bibitem{cpt}
M.~M.~Sheikh-Jabbari, \PRL {\bf 84}, 5265 (2000), hep-ph/0001167.

\bibitem{gws}
K.~Morita, hep-th/0011080.

\bibitem{HR}
J.~L.~Hewett, F.~J.~Petriello and T.~G.~Rizzo, hep-ph/0010354.

\bibitem{atom}
I.~Mocioiu, M.~Pospelov and R.~Roiban, \PLB {\bf B489}, 390 (2000),
hep-ph/0005191.

\bibitem{sheik}
I.~F.~Riad and M.~M.~Sheikh-Jabbari, \JHEP {\bf 08}, 045 (2000);
N.~Chair and M.~M.~Sheikh-Jabbari, hep-th/0009037.

\bibitem{lamp}
M.~Chaichian, M.~M.~Sheikh-Jabbari and A.~Tureanu, hep-th/0010175.

\bibitem{gold}
H.~D.~Abarbanel and M.~L.~Goldberger, Phys. Rev. {\bf 165}, 1594, (1968).

\bibitem{add}
N.~Arkani-Hamed, S.~Dimopoulos, and G.~Dvali, \PLB {\bf B429}, 263, (1998), 
and \PRD {\bf D59}, 086004, (1999); 
I.~Antoniadis, N.~Arkani-Hamed, S.~Dimopoulos, and G.~Dvali, \PLB {\bf B436}, 
257, (1998).

\bibitem{rs}
L.~Randall and R.~Sundrum, \PRL {\bf 83}, 3370, (1999), and 
{\it ibid}, 4690, (1999). 

\bibitem{ginz}
I.~F.~Ginzburg, G.~L.~Kotkin, V.~G.~Serbo and V.~I.~Telnov, Nucl. Instr.
Methods {\bf 205} 47 (1983); I.~F.~Ginzburg, G.~L.~Kotkin, S.~L.~Panfil,
V.~G.~Serbo and V.~I.~Telnov, Nucl. Instr. Methods {\bf 219} 5 (1984);
V.~I.~Telnov, Nucl. Instr. Methods {\bf A 294} 72 (1990).


\end{thebibliography}
\end{document}